\documentclass[prd,superscriptaddress,nofootinbib,twocolumn,showpacs,preprintnumbers]{revtex4}

\usepackage{amsfonts}
\usepackage{amsmath}
\usepackage{amssymb}
\usepackage{amsthm}
\usepackage{bm}
\usepackage{dcolumn}
\usepackage{epsfig}
\usepackage{graphicx}
\usepackage{graphics}
\usepackage[latin1]{inputenc}
\usepackage{latexsym}
\usepackage{rotating}
\usepackage[dvips]{color}

\newcommand\be{\begin{equation}}
\newcommand\ba{\begin{eqnarray}}
\newcommand\ee{\end{equation}}
\newcommand\ea{\end{eqnarray}}

\newcommand{\CS}{{\mbox{\tiny CS}}}
\newcommand{\GR}{{\mbox{\tiny GR}}}

\newcommand{\pont}{{\,^\ast\!}R\,R}

\begin{document}
\title {Double Binary Pulsar Test of Chern-Simons Modified Gravity}

\author{Nicol\'as Yunes}
\affiliation{Department of Physics, Princeton University, Princeton, NJ 08544, USA.}

\author{David N.~Spergel}
\affiliation{Princeton Center for Theoretical Science, Princeton University, Princeton, NJ 08544, USA.}
\affiliation{Department of Astrophysical Sciences, Princeton University, Princeton, NJ 08544, USA.}


\begin{abstract}

Chern-Simons modified gravity is a string-theory and loop-quantum-gravity inspired effective theory that modifies 
General Relativity by adding a parity-violating Pontryagin density to the Einstein-Hilbert action multiplied by a 
coupling scalar. We strongly constrain non-dynamical Chern-Simons modified gravity with 
a timelike Chern-Simons scalar through observations of the double binary pulsar PSR J$0737-3039$A/B.
We first calculate Chern-Simons corrections to the orbital evolution of binary systems. We find that 
the ratio of the correction to periastron precession to the general relativistic prediction scales 
quadratically with the semi-major axis and inversely with the square of the object's radius. Binary pulsar 
systems are thus ideal to test this theory, since periastron precession can be measured with sub-degree accuracies
and the semi-major axis is millions of times larger than the stellar radius. Using data from  
PSR J$0737-3039$A/B we dramatically constrain the non-dynamical Chern-Simons coupling to 
$M_{\CS} := 1/|\dot{\vartheta}| > 33 \; {\textrm{meV}}$, approximately a hundred billion times better 
than current Solar System tests.
	
\end{abstract}

\maketitle

\section{Introduction}
 
String theory is an intricate web of mathematically beautiful hypothesis that promises to unify all forces of nature. General Relativity (GR) is expected to be its low-energy limit with possible higher-order curvature corrections. To date, however, string theory remains intrinsically difficult to test experimentally, because these curvature-corrections are believed to be perturbatively Planck suppressed. Dynamical situations with large spacetime curvature could lead to non-linear couplings and enhance such curvature corrections to a constrainable realm. 

One such curvature correction is the {\emph{parity-violating}} Pontryagin density, which in addition to the Einstein-Hilbert term defines an effective theory: Chern-Simons (CS) modified gravity~\cite{jackiw:2003:cmo}. In four dimensions, this density is a topological term that does not contribute to the field equations, unless its coupling is non-constant or promoted to a scalar field~\cite{Smith:2007jm}. From a string theoretical standpoint, the Pontryagin correction is inescapable, if one is to have a mathematically consistent theory that is anomaly-free~\cite{Polchinski:1998rr,Alexander:2004xd,Svrcek:2006yi}. From an experimental standpoint, the search for the breakage of fundamental symmetries can provide hints that can guide theorist toward the correct ultraviolet completion of GR.   

CS modified gravity is also motivated from the standard model of elementary particles and from Loop
Quantum Gravity (LQG). For example, in particle physics, we know that an asymmetry in the number of left- and 
right-handed fermions forces the fermion number current to be anomalous, in analogy to the triangle 
anomaly~\cite{AlvarezGaume:1983ig}. This anomaly leads to the inclusion of the Pontryagin density in 
an effective fermionic action~\cite{Grumiller:2007rv}. Similarly, it has been recently found that LQG also 
leads to CS modified gravity when the Barbero-Immirzi parameter is promoted to a pseudo-scalar field 
in the presence of fermions~\cite{Taveras:2008yf,Mercuri:2009zt,Mercuri:2009vk,Mercuri:2009zi,Calcagni:2009xz}. 

The signature of CS modified gravity is the enhancement of gravitational parity asymmetry, which in particular leads to frame-dragging modifications~\cite{Alexander:2007qe,Alexander:2007zg,Alexander:2007vt}. In GR, the gravitomagnetic sector of the metric couples to the spin and the orbital angular momentum of gravitating systems, leading to corrections in their orbital evolution, such as precession of the orbital plane. In CS modified gravity, the gradient of the coupling scalar selects a preferred direction in spacetime that corrects this precession. Thus, observations of gravitomagnetic precession can be used to test the validity of the effective theory~\cite{Alexander:2007zg,Alexander:2007vt}. 

In the Solar System, this precession correction has already been studied for an externally prescribed (non-dynamical) CS coupling~\cite{Smith:2007jm}. Through comparisons with the LAGEOS and the Gravity Probe B experiment, bounds have been placed on the local magnitude of the time derivative of this field $\dot{\vartheta} \lesssim 10^{3} \; {\textrm{km}}$ or its associated {\emph{energy scale}} $M_{\CS} := 1/\dot{\vartheta} \gg 10^{-13} \; {\textrm{eV}}$ near Earth\footnote{When converting $\dot{\vartheta}$ into an energy scale we shall employ natural units. Otherwise, in this paper we use geometric units.}. 

From a theoretical standpoint, the effective mass scale for the CS term is uncertain.  While it could be as large as the Planck scale, it is intriguing to explore the possibility that the scale is around cosmological constant scale $ \Lambda^{1/4} \sim 1 \; {\textrm{meV}}$.
Such intrigue arises because the cosmological constant is an example of a quantity that according to string-theoretic predictions 
could be as much as $120$ orders of magnitude larger than the observed value, depending on the formulation. Since CS modified gravity
is also predicted by string theory, it is interesting to study whether its modifications are also observable at the cosmological constant scale.

The weakness of Solar System constraints can be qualitatively understood by focusing on the ratio of the CS precession correction to the GR expectation. For any binary system, this ratio scales as $({\cal{R}}_{ext}/{\cal{R}}_{ind})^{2}$, where ${\cal{R}}_{ext}$ and ${\cal{R}}_{ind}$ are the radius of curvature of the combined system and of either compact body respectively. For a binary system ${\cal{R}}_{ext} \sim a$, where $a$ is the semi-major axis, and ${\cal{R}}_{ind} \sim R$, where $R$ is the stellar radius. In the Solar System, $a = R_{+} + h$, where $h$ is the height to which satellites can be reliably placed in orbit, while $R=R_{+}$ is Earth's radius. Thus, the ratio ${\cal{R}}_{ext}/{\cal{R}}_{ind} -1  \sim h/R_{+} \ll 1$ and the CS effect is inherently small. For a binary pulsar, however, ${\cal{R}}_{ext}/{\cal{R}}_{ind} \sim {\cal{O}}(10^{5})$, which thus enhances the CS effect by a factor of approximately ${\cal{O}}(10^{10})$. 

In this paper, we study non-dynamical CS modified gravity in the far-field, applied to gravitomagnetic precession. 
We choose to work with the non-dynamical theory, since this has been studied in more detail 
(see eg.~\cite{alexander:2004:lfg,Konno:2008np,Cantcheff:2008qn,Alexander:2008wi,Alexander:2007vt,Alexander:2007zg,Smith:2007jm,Alexander:2007:gwp,Yunes:2008bu,Konno:2007ze,Tekin:2007rn,Guarrera:2007tu,Grumiller:2007rv,Yunes:2007ss}) 
and we choose the standard CS coupling scalar $\vartheta = \tau_{\CS} t$, where $\tau_{\CS}$ is a quantity we wish
to constrain with units of length. We recalculate the corrections to gravitomagnetic precession and solve the orbital perturbation equations to find the CS corrected, averaged rate of change of the periastron. Using the measurement of periastron precession from the double binary pulsar PSR J$0737-3039$A/B~\cite{Burgay:2003jj}, we place a bound on the magnitude of the time derivative of the CS coupling: $\dot{\vartheta} = \tau_{\CS} \lesssim 6 \times 10^{-9} \; {\textrm{km}}$ or equivalently 
$M_{\CS} \gtrsim 33 \; {\textrm{meV}}$ (in natural units), much stronger than the previous Solar System constraint. 

The remainder of this paper deals with the details of this calculation and it is divided as follows: Sec.~\ref{basics} defines CS modified gravity and presents the modified field equations of the theory; Sec.~\ref{astroph} tests the non-dynamical framework with the double binary pulsar; Sec.~\ref{conc} concludes and points to future research. 

We shall here employ the conventions in~\cite{Misner:1973cw}, with Greek letters ranging over spacetime indices, Latin letters over spatial indices only. We work exclusively in four spacetime dimensions, with the metric signature $(-,+,+,+)$. Round and square brackets in index lists denote symmetrization and
anti-symmetrization respectively, namely $T_{(\alpha \beta)}=\frac12
(T_{\alpha \beta}+T_{\beta \alpha})$ and $T_{[\alpha \beta]}=\frac12 (T_{\alpha \beta}-T_{\beta \alpha})$.
The Einstein summation convention is employed unless otherwise specified and geometric units ($G = c = 1$) are used mostly throughout the paper, 
except when relating our results to those of Ref.~\cite{Smith:2007jm}, where we use natural units ($h=c=1$). 

\section{Chern-Simons Modified Gravity and the Far Field Solution}
\label{basics}

Let us begin by summarizing the basic equations of CS modified gravity 
(see eg.~\cite{review} for a pedagogical review or~\cite{jackiw:2003:cmo,Smith:2007jm,Grumiller:2007rv} 
for more details). The CS action is here defined by
\be
\label{CSaction}
S = S_{\rm EH} + S_{\rm CS} +  S_{\vartheta} + S_{\rm mat},
\ee
where 
\ba
\label{EH-action}
S_{\rm{EH}} &=& \kappa \int_{{\cal{V}}} d^4x  \sqrt{-g}  R, 
\\
\label{CS-action}
S_{\rm{CS}} &=& \frac{\alpha}{4} \int_{{\cal{V}}} d^4x  \sqrt{-g} \; 
\vartheta \; \pont\,,
\\
\label{Theta-action}
S_{\vartheta} &=& - \frac{\beta }{2} \int_{{\cal{V}}} d^{4}x \sqrt{-g} \left[ g^{\alpha \beta}
\left(\nabla_{\alpha} \vartheta\right) \left(\nabla_{\beta} \vartheta\right) + 2 V(\vartheta) \right], \quad
\\
S_{\textrm{mat}} &=& \int_{{\cal{V}}} d^{4}x \sqrt{-g} {\cal{L}}_{\textrm{mat}}.
\ea
Equation~\ref{EH-action} is the Einstein-Hilbert action, Eq.~\eqref{CS-action} is the
CS correction, Eq.~\eqref{Theta-action} is the action for the scalar field and the last 
equation represents additional matter degrees of freedom. In these equations, 
$\kappa^{-1} = 16 \pi G$, $\alpha$ and $\beta$ are {\it{dimensional}} coupling constants,
 $g$ is the determinant of the metric, $\nabla_{\alpha}$ is the covariant derivative associated with the metric $g_{\alpha \beta}$, and $R$ is the Ricci scalar. The CS action depends on the Pontryagin density $\pont$, namely
\be
\label{pontryagindef}
\pont= R \tilde R = {\,^\ast\!}R^{\alpha}{}_{\beta}{}^{\gamma \delta} R^{\beta}{}_{\alpha \gamma \delta}\,,
\ee
where the dual Riemann-tensor is  
\be
\label{Rdual}
{^\ast}R^{\alpha}{}_{\beta}{}^{\gamma \delta}=\frac12 \epsilon^{\gamma \delta \epsilon \sigma}R^{\alpha}{}_{\beta \epsilon \sigma}\,,
\ee
and $\epsilon^{\gamma \delta \epsilon \sigma}$ the 4-dimensional Levi-Civita tensor. 

From the action we can derive the CS modified field equations~\cite{jackiw:2003:cmo,Smith:2007jm}
\ba
\label{EEs}
G^{}_{\mu \nu} + \frac{\alpha}{\kappa} C^{}_{\mu \nu}  &=& \frac{1}{2 \kappa} \left(T_{\mu \nu}^{\rm{mat}} + T_{\mu \nu}^{(\vartheta)}\right),
\\
\beta \square \vartheta &=& \beta \frac{dV}{d\vartheta} - \frac{\alpha}{4} \pont,
\label{theta-evol-eq1}
\ea
where $G_{\mu \nu}$ is the Einstein tensor and the C-tensor $C^{}_{\mu \nu}$ is defined via~\cite{Grumiller:2008ie}
\be
\label{C-def}
C^{\mu \nu} :=  v^{}_{\sigma}
  \epsilon^{\,\sigma \beta \alpha  (\mu} \nabla_{\alpha} R^{\nu)}{}_{\beta} +
v^{}_{\sigma \tau} {\,^\ast\!}R^{\sigma (\mu \nu)  \tau } \,. 
\ee
where $\square$ is the D'Alembertian operator, and $v_{\sigma} = \nabla_{\sigma} \vartheta$ and $v_{\sigma \alpha} = \nabla_{\sigma} \nabla_{\alpha} \vartheta$ are the covariant velocity and acceleration of the scalar field. The matter stress-energy tensor is $T_{\mu \nu}^{\rm{mat}}$ and the scalar-field stress-energy is $T_{\mu \nu}^{(\vartheta)}$ [see eg.~Eq.~$(67)$ in~\cite{Yunes:2007ss}], where 
\be
\label{Tab-theta}
T_{\alpha \beta}^{\vartheta} 
=   \beta  \left[  v_{\alpha} v_{\beta} - \frac{1}{2}  g_{\alpha \beta} v_{\sigma} v^{\sigma} -  g_{\alpha \beta}  V(\vartheta)  \right].
\ee
The scalar-field potential $V(\vartheta)$ depends on the fundamental theory from which CS modified gravity derives. In string theory, $\vartheta$ is a moduli field with a shift symmetry that forces the potential to vanish. One can generally
employ this assumption to set the potential to zero.

One of the main ingredients of CS modified gravity is the {\emph{CS coupling scalar}} $\vartheta=\vartheta(x^{\mu})$, which is a function of spacetime. If this field were a constant, then the modified theory reduces to GR because the Pontryagin density is a purely topological term. The coupling constant $\alpha$ determines the coupling strength of the CS scalar and the Riemann curvature, while the coupling constant $\beta$ determines the magnitude of the energy in the scalar field. The choice of units for either $(\alpha,\beta,\vartheta)$ fixes the units for the rest. For example, if $[\alpha] = L^{A}$ where $L$ is a unit of length and $A$ is any real number, 
then $[\vartheta] = L^{-A}$ and $[\beta] = L^{2 A - 2}$, where we required the action to be dimensionless when using natural units ($h = 1$). If instead we use geometric units ($G=c=1$), then the action has units of $L^{2}$, which if $[\vartheta] = L^{A}$, requires $[\alpha] = L^{2 - A}$ and $[\beta] = L^{-2A}$.

CS modified gravity can be classified into two distinct sub-theories: a dynamical one and non-dynamical one. In the dynamical theory, $\beta$ and $\alpha$ are arbitrary and the field equations are written in Eq.~\eqref{EEs} and~\eqref{theta-evol-eq1}. In the non-dynamical theory, $\beta = 0$ at the level of the action, and the field equations become 
\ba
\label{EEs-non-dyn}
G^{}_{\mu \nu} + \frac{\alpha}{\kappa} C^{}_{\mu \nu}  &=& \frac{1}{2 \kappa} T_{\mu \nu}^{\rm{mat}},
\\
\label{eq:constraint-nd}
0 &=& \pont.
\ea
The evolution equation for $\vartheta$ thus becomes a differential constraint, the so-called Pontryagin constraint, for the space of allowed solutions, while the scalar field is an externally prescribed quantity. 

In this paper, we shall choose to work with the non-dynamical theory\footnote{Henceforth, we choose $\alpha = \kappa$,
following~\cite{jackiw:2003:cmo}}, since this has already been studied
in detail and constraints (albeit weak) have already been placed on the strength of the 
correction~\cite{Alexander:2007qe,Alexander:2007zg,Alexander:2007vt,Smith:2007jm}.
In the non-dynamical framework, the functional form of the CS coupling scalar $\vartheta$ is not pre-determined.
When CS modified theory was originally proposed~\cite{jackiw:2003:cmo}, a specific choice was made, namely
$\vartheta_{c} = \tau_{\CS} t$, where $\tau_{\CS}$ is a constant with dimensions of length. 
A possible interpretation for $\vartheta_{c}$ is as an ``arrow of time''
since its associated embedding coordinate becomes $v_{\mu}^{c} = [\tau_{\CS},0,0,0]$. 
From a mathematical standpoint, this choice is convenient, since it leaves the CS action time-translation and
 reparameterization invariant (see eg.~\cite{jackiw:2003:cmo} for more details). From a physical standpoint, 
this choice is also convenient, since the Schwarzschild solution is automatically recovered for stationary and 
spherically symmetric backgrounds, and thus, most Solar System tests are automatically passed. The only
tests that are not automatically passed are those that involve gravitomagnetism, such as LAGEOS and Gravity Probe B,
and these experiments have been used to constraint $\dot{\vartheta}_{c} = \tau_{\CS} \lesssim 10^{3} \; {\textrm{km}}$~\cite{Smith:2007jm}. 
In Appendix~\ref{far-field}, we present some informal arguments for why $\vartheta = \vartheta_{c}$ might be the only
allowed functional for for the CS field in the Solar System, although a formal proof is still lacking.

With such a choice of $\vartheta$, one can solve the linearized field equations for the metric components. In 
Appendix~\ref{far-field}, we show explicitly that the temporal-temporal and spatial-spatial sectors of the 
modified field equations are automatically satisfied, which implies that scalar gravitational perturbations
are unaffected by the CS modification. The $0i$ field equations, however, are CS corrected, 
but they can be solved to linear order in $\tau_{\CS}$ via 
(see Appendix~\ref{far-field} for more details)
\be
g_{0k} =  - 4  \int \frac{v_{k} \rho'}{|x - x'|} d^{3}x' 
- 2 \tau_{\CS}  \int d^{3}x' \frac{(\vec\nabla\rho \times \vec{v})_{k}}{|x - x'|} .
\label{frame-drag-pot}
\ee
where $\rho$ is a matter density distribution, while $v_{i}$ is its three-velocity,  
$\times$ is the Euclidean cross product and we have neglected any time dependance of the Newtonian 
gravitational potential. The vectorial solution presented here is similar to that found 
in~\cite{Alexander:2007zg,Alexander:2007vt}, except that here we consider generic density distributions. 
One can show that in the limit as $\rho \to m \;\delta^{3}(x^{i})$, Eq.~\eqref{frame-drag-pot} reduces 
identically to Eq.~$(44)$ in~\cite{Alexander:2007zg}, with the appropriate choice of $\tau_{\CS}$. 

The gravitomagnetic potential presented in Eq.~\eqref{frame-drag-pot} is similar to that found in~\cite{Smith:2007jm}. 
One can show that if $\rho v_{i} $ is replaced with the stress-energy component appropriate to a homogeneous rotating sphere, 
then this potential reduces to Eq.~$(B4)$ in~\cite{Smith:2007jm} with the appropriate choice of $\tau_{\CS}$ and to 
linear order. Care must be taken, however, when solving explicitly Eq.~\eqref{frame-drag-pot} with a non-trivial density
distribution, as boundary terms might be required to ensure the junction conditions are satisfied~\cite{Smith:2007jm}.  

\section{Astrophysical Tests of CS Modified Gravity}
\label{astroph}

\subsection{Weak-Field Tests}

The lack of a CS correction to the scalar sector of the gravitational perturbations implies that most astrophysical process are unaffected. For example, the equations of structure formation remain untouched because the Poisson equation is not CS corrected and the stress-energy tensor remains locally conserved. The vectorial sector of the metric, however, is CS modified in a normal direction relative to the GR prediction. For randomly oriented velocities, the average value of the leading order 
CS correction (that shown in Eq.~\eqref{frame-drag-pot}) in fact identically vanishes,  simply because the correction is odd in $v^{i}$ and $<v^{i}> = 0$. In many astrophysical scenarios, however, the velocity field is not randomly oriented. One such case are binary systems, where the CS correction leads to an anomalous frame-dragging effect. 

Anomalous frame-dragging induces modifications on a variety of astrophysical processes, such as the formation of accretion discs around protoplanetary systems and the evolution of neutron star spins. The CS correction, however, would be hard to detect in such processes because it scales inversely with the radius of curvature of the system, as one can see from Eq.~\eqref{frame-drag-pot}. Galactic radii are on the kpcs scale, which renders the ratio of the CS correction to the GR prediction on the ${\cal{O}}(10^{-17})$ if we saturate the Solar System constraint~\cite{Smith:2007jm}.

Although the CS correction is insignificant in the evolution of non-compact astrophysical sources, this is not the case for binary systems. For example, inspiraling black hole binaries would be ideal laboratories to test CS modified gravity. Such systems do not radiate electromagnetically unless surrounded by an accretion disk, but gravitational wave observations with space-based or earth-based detectors could be used to test the modified theory~\cite{Alexander:2007:gwp,Yunes:2008bu,Yunes:2009hc,Sopuerta:2009iy}. Such observations would be sensitive to the integrated history of the CS term, instead of its instantaneous value.

Another type of binary system that shall be used in this paper to test the modified theory are binary pulsars. In such systems, there are 
two important scales the CS correction could couple to: the radius of curvature of the system, which is proportional to the semi-major 
axis $a$; and the radius of curvature of either component, which is proportional to the radius of either body $R$. As we shall see, the 
GR prediction for the precession of the periastron scales as $a^{-3}$, while the CS correction scales as $a^{-1} R^{-2}$, which implies 
that observed binary systems are preferred laboratories to test the modified theory since $a/R \sim {\cal{O}}(10^{5})$.

One might worry at this juncture that the CS correction dominates over the GR solution 
when $M/a \ll 1$, since the former seems to decay more slowly with semi-major axis than the latter. 
In this paper, however, we shall assume that the solution in Eq.~\eqref{frame-drag-pot} applies when 
the CS correction is small relative to the GR solution. This is indeed the case provided $\tau_{\CS} \ll R^{3}/a^{2}$,
 which for the binary system under consideration becomes  $\tau_{\CS} \ll 10^{-6}$ meters. 
 We shall see that the bound derived in this paper forces $\tau_{\CS}$ to be well below this value, 
and thus, the approximation made are indeed valid. 

\subsection{Binary Pulsar Test}

Consider binary systems of spinning neutron stars, whose orbital evolution we shall model through a geodesic study of a compact object in the background of a rotating, homogeneous sphere. Following~\cite{Smith:2007jm}, the stress-energy tensor of this sphere will be described by $T_{0i} = -j_{i}$, where the current $j_{i} = \rho_{0} \left( \vec{\Omega} \times \vec{r} \right)_{i} \Theta(R-r)$, with $\rho_{0}$ some constant density, $\Omega^{i} = [0,0,\Omega]$ the angular velocity vector in Cartesian coordinates, $r^{i} = [x,y,z]$ the distance from the center of the sphere to a field point and $R$ the radius of the sphere. The total mass of this sphere is $M = 3 \rho_{0}/(4 \pi R^{3})$, while its spin angular momentum $J^{i} = I \Omega^{i}$, where $I = (2/5) M R^{2}$ is the moment of inertia

The motion of the compact body in this background is governed by the geodesic equations $\vec{a} = -4 \vec{v} \times \vec{B}$, where we have neglected time-dependent scalar potentials and where $\vec{a}$ and $\vec{v}$ are the three-acceleration and three-velocity of the compact object. The  gravitomagnetic field is $\vec{B} := \vec{\nabla} \times \vec{A}$, where the gravitomagnetic potential is $A_{i} := - g_{0i}/4$. As anticipated in the previous section, both the field and the potential have been computed for this stress-energy to arbitrary order in $\tau_{\CS}$~\cite{Smith:2007jm}, but we shall here work only to leading order $\tau_{\CS}$, since as we shall see second-order effects will be negligible. The gravitomagnetic field can then be written as $\vec{B}_{\CS} = \vec{B} - \vec{B}_{\GR}$, with
\be
\label{BCS}
\vec{B}_{\CS} = 
\frac{c_{0}}{r} \cos[\xi(r)]  \left[ 
 \vec{{\cal{J}}}
-\tan\xi \left(\vec{{\cal{J}}} \times \hat{r}\right) 
- \left(\vec{{\cal{J}}} \cdot \hat{r}\right)  \hat{r} \right],
\ee
where $\xi(r) = 2 r /\tau_{\CS}$,  $c_{0} = 15 \tau_{\CS}/(4 R)  \sin[\xi(R)]$, $\hat{r} = \vec{r}/r$, $\vec{{\cal{J}}} = \vec{J}/R^{2}$ and $\cdot$ the Euclidean dot product. 
Note that Eq.~\eqref{BCS} accounts both for the homogeneous solution of~\cite{Alexander:2007zg,Alexander:2007vt} and the boundary term found in~\cite{Smith:2007jm}.

From the gravitomagnetic field, we can straightforwardly compute the CS correction to the geodesic acceleration by taking the cross-product with the velocity vector
\ba
\vec{a}_{\CS} &=& - \frac{4 c_{0}}{r}
\left\{ \cos[\xi(r)] \left(\vec{v} \times \vec{{\cal{J}}} \right) 
\right. 
\nonumber \\
&-&\left.
\sin[\xi(r)]  \left[ \vec{{\cal{J}}} \left(\vec{v} \cdot \hat{r} \right) - \hat{r} \left(\vec{{\cal{J}}} \cdot \vec{v} \right) \right] 
\right. 
\nonumber \\
&-&\left.
\cos[\xi(r)] \left(\vec{{\cal{J}}} \cdot \hat{r} \right) \left(\vec{v} \times \hat{r} \right) \right\},
\ea
plus subdominant terms that scale with higher powers of the derivatives of the CS scalar. 

One must be careful when expanding solutions in $\tau_{\CS}$, since this quantity has units of length, and thus, corrections will arise as combinations of ${\cal{O}}(\tau_{\CS}/R)$ and ${\cal{O}}(\tau_{\CS}/a)$. The approximations made so far hold provided these combinations are much smaller than unity. 
As we already argued, $\tau_{\CS} \ll R^{3}/a^{2}$, which supersedes the above requirements. Care must be taken, however, since the argument of the oscillatory functions scales as $1/\tau_{\CS}$, and thus, any spatial derivatives of $A_{i}$ (or $g_{0i}$) will be larger than $A_{i}$ (or $g_{0i}$) by one power of $\tau_{\CS}$. When this fact is taken into account, the results for the gravitomagnetic field found in~\cite{Alexander:2007zg,Alexander:2007vt} and~\cite{Smith:2007jm} are in formal agreement\footnote{As already discussed,~\cite{Smith:2007jm} obtains a boundary contribution that is not modeled in~\cite{Alexander:2007zg,Alexander:2007vt}, because the latter employed a point-particle approximation, while the former dealt with extended bodies.}.  

We shall here parameterize the trajectory of the compact object in terms of equatorial coordinates. We shall thus define the triad:
\ba
\hat{r} &=& \left[ \cos{u}, \cos{\iota} \sin{u}, \sin{\iota} \sin{u} \right],
\nonumber \\
\hat{t} &=& \left[ -\sin{u},\cos{\iota} \cos{u}, \sin{\iota} \cos{u} \right],
\nonumber \\
\hat{n} &=& \left[0,-\sin{\iota},\cos{i}\right],
\ea
to describe radial, transverse and normal directions relative to the comoving frame in the orbital plane. The quantity $\iota$ is the inclination angle, $w$ is the argument of periastron, $u = f + w$ with $f$ the true anomaly and $\Omega=0$ is the right ascension of the ascending node, chosen in this way so that the line of nodes is co-aligned with the $\hat{x}$ vector~\cite{satbook}. 

The perturbation equations for the variation of the Keplerian orbital elements is governed by the projection of the geodesic acceleration on this triad. To leading order in $\tau_{\CS}$, however, only the radial projection $a_{r} := \vec{a} \cdot \hat{r}$ is CS modified, leading to $a_{r}^{\CS} := a_{r} - a_{r}^{\GR}$: 
\be
a_{r}^{\CS} = -4 c_{0} \dot{u}  {\cal{J}} \left\{ \cos{\iota}  \cos\left[\xi(r)\right]  + \sin{\iota} \cos{u} \sin\left[\xi(r)\right]  \right\}.
\ee

We can now compute the variation of the orbital elements by studying the perturbation equations~\cite{satbook}
\ba
\frac{da}{dt} &=& \frac{2}{n \sqrt{1 - e^{2}}} \left[ e \; a_{r} \sin{f} + a_{t} \frac{p}{r} \right],
\nonumber \\
\frac{de}{dt} &=& \frac{\sqrt{1 - e^{2}}}{na} \left\{ a_{r} \sin{f} + a_{t} \left[ \cos{f} + \frac{1}{e} \left(1 - \frac{r}{a} \right) \right] \right\},
\nonumber \\
\frac{di}{dt} &=& \frac{1}{n \; a \sqrt{1 - e^{2}}} a_{n} \frac{r}{a} \cos{u},
\nonumber \\
\frac{d\Omega}{dt} &=& \frac{1}{n \; a \sin{\iota} \sqrt{1 - e^{2}}} a_{n} \frac{r}{a} \sin{u},
\nonumber \\
\frac{d\omega}{dt} &=& \frac{\sqrt{1 - e^{2}}}{n\;a\;e} \left[-a_{r} \cos{f} + a_{t} \left(1 + \frac{r}{p} \right) \sin{f} \right] 
\nonumber \\
&-& \cos{\iota} \frac{d\Omega}{dt},
\nonumber \\
\frac{d{\cal{M}}}{dt} &=& n - \frac{2}{n\;a} a_{r} \frac{r}{a} - \sqrt{1 - e^{2}} \left(\frac{d\omega}{dt} + \cos{\iota} \frac{d\Omega}{dt} \right).
\ea
where $n = \sqrt{M/a^{3}}$ is the unperturbed Keplerian mean motion, $e$ is the eccentricity, and ${\cal{M}}$ is the mean anomaly. Since $\vec{a}^{\CS} \cdot \hat{t} = {\cal{O}}(\tau_{\CS}^{2}) = \vec{a}^{\CS} \cdot \hat{n}$, $\dot\Omega_{\CS} = 0$
while $\dot{w}_{\CS} = - a_{r}/(n\; a \; e)  \cos{f}$, to leading order in the eccentricity.

The average of the rate of change of $w$ can be computed by integrating $\dot{\omega}$ over one orbital period:
\be
<\dot{\omega}> := \int_{0}^{T} \frac{\dot{\omega}}{P} dt =  \int_{0}^{2 \pi} \dot{\omega} \frac{\left(1 - e^{2}\right)^{3/2}}{2 \pi \left(1 + e \cos{f} \right)^{2}} df,
\label{ave}
\ee
during which we shall assume the pericenter is approximately constant, so that $\dot{u} \sim \dot{f} =  n \left(1 + e \cos{f} \right)^{2} \left(1 - e^{2}\right)^{-3/2}$, and the motion of the compact object can be described by a Keplerian ellipse, where
$r = a \left(1 - e^{2} \right) \left(1 + e \cos{f}\right)^{-1}$. This last assumption is justified by the fact that the motion of test particles about any arbitrary background remains unchanged relative to the GR prediction, ie.~test-particle motion satisfies the geodesic equation both in GR and in CS modified gravity. Such is the case provided the
strong-equivalence principle is satisfied, which is guaranteed by the Pontryagin constraint in the non-dynamical version of the theory or by the scalar field equation
of motion in the dynamical version (see~\cite{Sopuerta:2009iy} for a proof). Finally, the integrals in Eq.~\eqref{ave} shall be approximated with a small eccentricity expansion $e \ll 1$. 

The averaged rate of change of the periastron can then be decomposed into a GR prediction plus a CS correction, where the latter is given by 
\be
\left< \dot{w} \right>_{\CS} 
= \frac{15}{2a^{2} e} \frac{J}{R^{2}} \frac{\tau_{\CS}}{R}  X \sin\left( \frac{2 R}{\tau_{\CS}} \right) \sin\left(\frac{2 a}{\tau_{\CS}}\right),
\label{dwdt_ave}
\ee
and where the projected semi-major axis is $X := a \sin{\iota}$. Equation~\eqref{dwdt_ave} neglects terms of order unity, since the dominant contribution scales as $e^{-1}$ and we here concentrate on systems with small but non-vanishing eccentricity. The $e^{-1}$ scaling occurs because although $\dot{w}$ scales as $\cos{f}$, so does $a_{r}^{\CS}$, and thus, the leading order term in $e$ does not vanish upon integration, unlike in the GR case. The orbital orientation, however, is ill-defined for exactly zero eccentricity, and thus, the limit $e \to 0$ is meaningless. The scaling in the precession of the periastron of Eq.~\eqref{dwdt_ave} is consistent with other precession results studied in the Solar System ~\cite{Smith:2007jm}. As discussed in the Introduction, the ratio of CS correction to the GR expectation scales as $a^{2} \tau_{\CS}/R^{3}$, since $\left<\dot{\omega}\right>_{\GR} \sim J/a^{3}$. In the Solar System, however, $a/R$ is very close to unity, while for binaries $a/R \sim {\cal{O}}(10^{6})$.

At this junction, one might worry that the calculation of the CS modification to $\dot{w}$ is not sufficient to place a bound on the non-dynamical theory with the binary pulsar, since other relevant quantities that play an important role in the test might also be CS modified~\cite{lrr-2006-3}. Although this is generally true, the CS modification to other quantities turns out to be either identically zero or subleading. This is so because $C_{00}$ identically vanish for $\vartheta \propto t$ in the non-dynamical formalism, while $C_{ij}$ is proportional to gravitational wave perturbations only. These waves will be CS modified, but the CS correction to the quadrupole formula is subleading [$h_{ij}^{\CS} \propto (\tau_{\CS}/D_{L}) d^{3} I_{ij}/dt^{3}$, where $I_{ij}$ is the quadrupole moment and $D_{L}$ is the luminosity distance to the source~\cite{Sopuerta:2009iy}], and thus, the CS correction to the rate of change of the binary period $\dot{P}$ is also subleading. Since the only post-Keplerian parameters that is CS corrected to leading order is $\dot{\omega}$, we can treat, as a first approximation, all other orbital quantities as given in~\cite{Kramer:2006nb}. 

With these remarks in mind, observations of the precession of the periastron in the double binary pulsar PSR J$0737-3039$A/B~\cite{Burgay:2003jj} can be used to test CS modified gravity. We shall treat pulsar $A$ as the rotating homogeneous sphere, and pulsar B as the test body in orbit around the sphere, where the bodies are sufficiently separated that we can neglect tidal interactions. The relevant system parameters are~\cite{Kramer:2006nb}
the mass $M_{A} = M \approx 1.34 M_{\odot}$,
the projected semi-major axis $X \approx 1.41 \; {\textrm{s}}$,
the eccentricity $e_{b} \approx 0.088$
and the inclination angle $\iota \approx 89 (-76,+50) \deg$.
From the projected semi-major axis we can deduce that $a_{b} \approx 4.24 \times 10^{5} \; {\textrm{km}}$, where we used the nominal value for the inclination angle. We assume the standard value for the moment of inertia of body A $I \approx  10^{38} \; {\textrm{kg}} \; {\textrm{m}}^{2}$, which leads to a radius of $R_{A} \approx  9.69 \; {\textrm{km}}$ and an angular momentum of $J_{A} \approx  2.8 \times 10^{40} \; {\textrm{kg}} \; {\textrm{m}}^{2} \; {\textrm{s}}^{-1}$, since
the rotational period has been determined to be $22$ milliseconds~\cite{Kramer:2006nb,Iorio:2008xz}. The precession of the periastron
has been measured to be $\dot{\omega} \sim 16.96$ degrees per year, in complete agreement with the GR expectation, 
with an overall uncertainty of approximately $\delta = 0.05$ degrees per year\footnote{The precession of the periastron has been measured to higher accuracy for 
pulsar A, but we adopt here the larger uncertainty so as to derive a conservative bound on $m_{\CS}$.}~\cite{Kramer:2006nb,Iorio:2008xz}. 

\begin{figure}
\includegraphics[scale=0.33,clip=true]{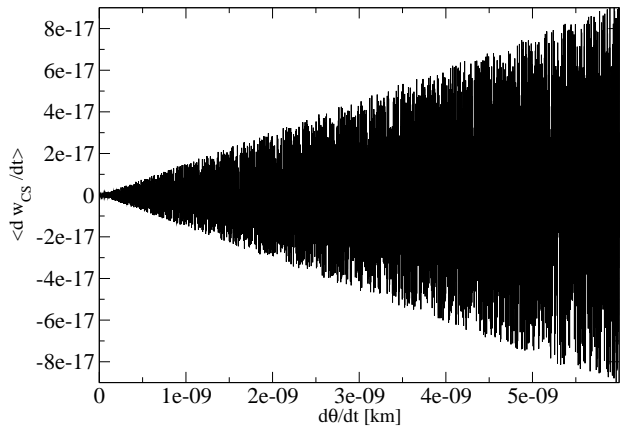}
\caption{\label{fig} CS correction to the precession of the periastron as a function of $\tau_{\CS}$ for the system parameters of PSR J$0737-3039$A/B.}
\end{figure}

We can then constraint $\tau_{\CS}$ by requiring graphically  that $\left<\dot{\omega}_{\CS}\right>$ be less than $\delta$. The uncertainty in geometric units becomes 
$\delta = 9.2 \times 10^{-17} {\textrm{radians}} \; {\textrm{km}}^{-1}$.
Figure~\ref{fig} presents a plot of $\left<\dot{\omega}_{\CS}\right>$  as a function of $\tau_{\CS}$. 
A $1 \sigma$ constraint can then be derived from this figure, namely 
$\dot{\vartheta} = \tau_{\CS} \lesssim 6 \times 10^{-9} \; {\textrm{km}}$, 
or simply 
$M_{\CS} := \dot{\vartheta}^{-1} = \tau_{\CS}^{-1} \gtrsim 33 \; {\textrm{meV}}$ 
in natural units, which is approximately $10^{11}$ times stronger than current Solar System constraints. 

We have checked that terms higher order in $e$ or $\tau_{\CS}$ do not significantly affect this bound, which is however primarily affected by uncertainties in the semi-major axis. Even with the most pessimistic choice of $a$, the bound deteriorates only by a factor of twenty, still leading to a constraint $10^{10}$ times stronger than the Solar System one. Also note that this bound is consistent with the approximation made to derive the CS correction to periastron precession.

\section{Conclusions}
\label{conc}

We have studied non-dynamical CS modified gravity with a time-like CS coupling scalar. Until now, the only constraint on non-dynamical 
CS modified gravity ($M_{\CS} \lesssim 10^{-13} \; {\textrm{eV}}$) came from
Solar System experiments due to CS corrections to frame-dragging~\cite{Smith:2007jm,Alexander:2007zg,Alexander:2007vt}. We here calculated 
the leading-order CS correction to post-Keplerian parameters of binary systems. We find that the precession of the periastron is the only parameter
that is CS corrected to leading order. This corrections is such that its ratio to the GR expectation scales as $a^{2}/R^{2}$, 
where $a$ is the semi-major axis and $R$ is the neutron star radius. For the binary pulsar considered here, this ratio is of 
${\cal{O}}(10^{10})$ which leads to an enormous enhancement over previous Solar System constraints: $M_{\CS} > 33 \; {\textrm{meV}}$. 
This constraint is approximately a hundred billion times stronger than current Solar System constraints.

Although this paper constrains the non-dynamical framework of CS modified gravity to unprecedented levels, it cannot do the same for the dynamical 
formulation. Corrections to post-Keplerian parameters in the dynamical theory are of high post-Newtonian order, because the Pontryagin density
vanishes to leading order. Meaningful tests of the dynamical formulation would then have to rely on strongly gravitating sources.

One such scenario is the inspiral and merger of compact objects. Dynamical CS modified gravity should correct both the trajectories of such objects 
as well as the generation of gravitational waves. A detection of such waves with LIGO or LISA could then be used to place stringent bounds on the 
dynamical formulation~\cite{Alexander:2007:gwp,Yunes:2008bu,Yunes:2009hc,Sopuerta:2009iy}. A program that pursues just such a calculation is 
currently ongoing~\cite{Yunes:2009hc,Sopuerta:2009iy}.   

\acknowledgments

We wish to thank Joe Taylor, Juan Maldacena, Richard O'Shaughnessy and Frans Pretorius for useful discussions and comments. N.~Y.~ acknowledges support from the NSF grant PHY-0745779, while D.~S.~acknowledges support from the NASA theory grant NNX08AH30G. 

\appendix
\section{Far-Field Solutions in Non-Dynamical CS Gravity}
\label{far-field}

In this Appendix, we study solutions to the modified field equations in the far-field. 
We begin by describing the approximation schemes employed and then tackle the
modified field equations order by order. We shall here initially allow $\vartheta$ to 
be arbitrary, but we shall present some informal arguments that suggest $\vartheta_{c}$ might
be the only solution allowed in the Solar System, albeit a formal proof is lacking.

\subsection{Approximation Schemes}
\label{small-coup}

Consider the far-field expansion of the line element
\be
ds^{2} = - ( 1 + 2 \phi) dt^{2}   + 2 g_{i} dt dx^{i} + ( 1 - 2 \psi) \delta_{ij} dx^{i} dx^{j},
\label{metric}
\ee
where $t,x^{i}$ are {\emph{Cartesian}} coordinates, $\delta_{ij}$ is the Euclidean metric, $(\psi,\phi)$ and $g_{i} := g_{0i}$ are scalar and vectorial perturbation potentials respectively in the longitudinal gauge ($\partial_{i} g^{i} = 0$).

The perturbation potentials and the matter sources that generate them shall be treated perturbatively, in a post-Newtonian (PN) sense, where the latter are assumed slowly-moving ($\epsilon := v/c \ll 1$), weakly-gravitating and isolated (see eg.~\cite{lrr-2006-3}). We shall ignore gravitational wave perturbations, since these have been partially studied elsewhere~\cite{jackiw:2003:cmo,alexander:2004:lfg,alexander:2005:bgw,Alexander:2007:gwp}. In particular, we shall model these sources via a perfect fluid stress energy tensor, such that $T_{00} = \rho = {\cal{O}}(\epsilon^{2})$, $T_{0i} = - \rho v_{i} = {\cal{O}}(v^{3})$ and $T_{ij} = {\cal{O}}(\epsilon^{4})$, where $\rho$ is density and $v^{i}$ is the three-velocity of the fluid. 

We shall be concerned here with binary systems, whose {\emph{exterior}} gravitational field (or metric) can be modeled in GR (to leading post-Newtonian order) via
\ba
\psi_{\GR} &\sim& \phi_{\GR} \sim \frac{m_{1}}{r_{1}}+ \frac{m_{2}}{r_{2}},
\nonumber \\
g^{i}_{\GR} &\sim& \frac{m_{1}}{r_{1}} v_{1}^{i}(t) + \frac{m_{2}}{r_{2}} v_{2}^{i}(t),
\label{GR-sol}
\ea
where $m_{1,2}$ are the component masses, $r_{1,2} = |x^{i} - x_{1,2}^{i}(t)|$ are their field point positions, with $x_{1,2}^{i}(t)$ the trajectories
and $v_{1,2}^{i}(t) = \dot{x}_{1,2}^{i}(t)$ the velocities. Note that these potentials are {\emph{not spherically-symmetric}} and are here expressed in Cartesian coordinates, 
in spite of the appearance of fiducial radial distances $r_{1,2}$. Such potentials become spherically symmetric only in the limit $m_{2} \to 0$ and $m_{1}$ fixed (or viceversa),
in which case one recovers a boosted, Schwarzschild metric in harmonic coordinates. If these objects are spinning, 
then the gravitomagnetic sector of the metric $g^{i}_{\GR}$ acquires more terms proportional to the spin angular momentum (see eg.~\cite{Alexander:2007zg,Alexander:2007vt}).
These equations can be derived by assuming a point-particle approximation, but we relegate any such 
details to the post-Newtonian reviews in~\cite{lrr-2006-3,lrr-2006-4}.

We shall concentrate on a rather special, yet physically reasonable subset of metric perturbations: {\emph{potentials that are small CS deformations of GR solutions}}. In other words, on top of the PN perturbative scheme, we shall employ a {\emph{small-coupling approximation}}. In the dynamical formalism,
this can be achieved by expanding in $\zeta := \alpha^{2}/(\kappa \beta M^{4}) \ll 1$, where $M$ is the gravitational mass (a length-scale) 
associated with $\rho$. In the non-dynamical scheme, the perturbation parameter becomes $\zeta = |\partial_{\mu} \vartheta|/M \ll 1$, 
which for $\vartheta = \vartheta_{c}$ becomes $\dot{\vartheta}/M = \tau_{\CS}/M \ll 1$. 

The combined use of a PN expansion and the small-coupling approximation defines a bivariate perturbation scheme, where both $\epsilon$ and $\zeta$ can be treated as independently small parameters. Moreover, in the dynamical framework, this scheme defines a boot-strapping framework in which one can first solve the evolution equation for $\vartheta$ in the non-CS corrected background, and then use this $\vartheta$ to solve the modified field equations to first-order in the CS correction. For more details on this boot-strapping scheme or the small-coupling approximation as applied to the dynamical theory, we refer the reader to~\cite{Yunes:2009hc}. 

Based on these considerations, we shall commonly make the decomposition $A = A^{GR} + \chi_{(A)}$, where $A$ is any metric perturbation, $A^{GR}$ is a GR solution of ${\cal{O}}(\zeta^{0})$ and $\chi_{A}$ is some undetermined potential of ${\cal{O}}(\zeta)$. Moreover, we shall require that $A$ and $\chi_{A}$ be at least of the same order in $\epsilon$, such that we can search for CS-deformed solutions. The metric perturbations shall then be expanded as
\be
\psi = \psi_{\GR} + \chi_{(\psi)},
\quad
\phi = \phi_{\GR} + \chi_{(\phi)},
\quad
g^{i} = g^{i}_{\GR} + \chi^{i}_{(g)}
\label{decomp}
\ee
where $\psi_{GR}$ and $\phi_{GR}$ are both of ${\cal{O}}(\epsilon^{2},\zeta^{0})$, $g^{i}_{\GR}$ is of ${\cal{O}}(\epsilon^{3},\zeta^{0})$, $\chi_{(\psi)}$ and $\chi_{(\phi)}$ are both of ${\cal{O}}(\epsilon^{2},\zeta)$, and $\chi^{i}_{(g)}$ is of ${\cal{O}}(\epsilon^{3},\zeta)$. The notation ${\cal{O}}(\epsilon^{m},\zeta^{n})$ stands for a term of 
${\cal{O}}(\epsilon^{m})$ or ${\cal{O}}(\zeta^{n})$. 

Such a decomposition neglects CS corrections that modify the leading-order behavior of GR. For example, we shall not consider a perturbation $\chi_{(\phi)}$ of ${\cal{O}}(\epsilon,\zeta)$. Such an assumption is justified on the basis of the small-coupling approximation and the fact that GR has been found to agree to incredible precision with Solar System experiments~\cite{lrr-2006-3}. The above expansion then guarantees that to lowest order all such experiments are passed, with the new potentials $\chi_{(\psi),(\phi),(g)}$ leading only to small perturbations.

\subsection{Scalar Metric Perturbations}

The scalar metric perturbations are traditionally solved for by studying the field equations of a theory 
to ${\cal{O}}(\epsilon^{2})$. For an arbitrary $\vartheta$, the modified field equations become
\ba
\label{00}
&& \nabla^{2} \psi = 4 \pi \rho,
\\
\label{0i}
&& \tilde{\epsilon}_{i}{}^{jk} \left[ \left(\partial_{j} \vartheta \right) \partial_{k} \nabla^{2} + \left( \partial_{jm} \vartheta \right) \partial_{mk} \right] \left( \psi + \phi\right) = 0,
\\
&& \left( \delta_{ij} \nabla^{2} - \partial_{ij} \right) \left( \phi - \psi \right) = 
- \left(\partial_{l} \dot\vartheta\right) \tilde{\epsilon}_{(i}{}^{kl} \partial_{j)k} \left( \psi + \phi \right), \quad \;
\label{ij}
\ea
where $\nabla^{2}$ is the Laplacian in Cartesian coordinates, $\epsilon^{ijk} := \epsilon^{0ijk}$ and $\tilde{\epsilon}^{ijk}$ is the Levi-Civita symbol.

The decomposition of Eq.~\eqref{decomp} simplifies the modified field equations. To ${\cal{O}}(\epsilon^{2},\zeta^{0})$, the field equations reduce
to those of GR, namely
\ba
\label{00-simped}
&& \nabla^{2} \psi_{\GR} = 4 \pi \rho,
\\
&& \left( \delta_{ij} \nabla^{2} - \partial_{ij} \right) \left( \phi_{\GR} - \psi_{\GR} \right) = 0,
\label{ij-simped}
\ea
where the temporal-spatial component automatically vanishes. Equation~\eqref{00-simped} leads to the standard GR 
solution~\cite{Smith:2007jm,Alexander:2007zg,Alexander:2007vt}
\be
\psi_{\GR} \propto- \int \frac{\rho(x')}{|x - x'|} d^{3}x',
\ee 
while Eq.~\eqref{ij-simped} implies $\phi_{\GR} = \psi_{\GR}$. 

To next order [${\cal{O}}(\epsilon^{2},\zeta)$], the modified field equations become
\ba
\label{00-O1}
&& \nabla^{2} \chi_{(\psi)} = 0,
\\
&& \tilde{\epsilon}_{i}{}^{jk} \left[ \left(\partial_{j} \vartheta \right) \partial_{k} \nabla^{2} + \left( \partial_{jm} \vartheta \right) \partial_{mk} \right] \psi_{\GR} = 0,
 \label{0i-scal}
\\
&& \left( \delta_{ij} \nabla^{2} - \partial_{ij} \right) \left( \chi_{(\phi)} - \chi_{(\psi)} \right) =
-2 \left(\partial_{l} \dot\vartheta\right) \tilde{\epsilon}_{(i}{}^{kl} \partial_{j)k}  \psi_{\GR}, \qquad \quad 
\label{ij-O1}
\ea
Equation~\eqref{00-O1} forces $\chi_{(\psi)}$ to be a solution to the Laplace equation. The remaining two equations [Eqs.~\eqref{0i-scal} and~\eqref{ij-O1}] 
are clearly satisfied if $\vartheta = \vartheta_{c}$. 

We have searched for other choices for $\vartheta$ that satisfy these equations and we have found the following sufficient conditions: 
\be
{\textrm{\bf{(i)}}} \quad \partial_{i} \vartheta =0, \qquad
{\textrm{\bf{(ii)}}} \quad \partial_{i} \vartheta = {\cal{O}}(\epsilon). 
\label{suff-cond}
\ee
Clearly, the choice $\vartheta = \vartheta_{c}$ satisfies either (i) or (ii). Option (i) forces both terms in Eq.~\eqref{0i-scal} to exactly vanish, as well as
the right-hand side of Eq.~\eqref{ij-O1}, while option (ii) forces these terms to vanish only perturbatively, 
since then $(\partial \vartheta) (\partial \psi_{\GR}) = {\cal{O}}(\epsilon^{3},\zeta)$. 
Either option then forces $\chi_{(\psi)} = \chi_{(\phi)}$, but since these functions must satisfy the Laplace 
equation and the metric (and thus $\chi_{(\psi),(\phi)}$) 
must be asymptotically flat at spatial infinity\footnote{Since the metric is asymptotically flat, it must decay as $1/r_{1}$ or $1/r_{2}$ 
at spatial infinity, but the Laplacian of these functions is non-vanishing. For example, in the limit of vanishing $m_{2}$, 
the Laplacian of $1/r$ becomes proportional to the Dirac delta function.}, 
we choose $\chi_{(\psi)} = 0 = \chi_{(\phi)}$.  

Although the conditions presented above are sufficient, we cannot formally prove that they are necessary
to satisfy the modified field equations. In other words, although we have failed
to find a solution to the above system of differential equations, we cannot prove that a solution does not necessarily exist. 

\subsection{Vectorial Metric Perturbations}

The vectorial sector of the metric perturbations can be solved for by studying the field equations to ${\cal{O}}(\epsilon^{3})$. 
For an arbitrary scalar field, these equations become
\ba
&& \tilde{\epsilon}^{ijk} \left(\partial_{i} \vartheta \right) \nabla^{2} \partial_{j} g_{k} = - \tilde{\epsilon}^{ijk}  \left(\partial_{il} \vartheta \right) \partial_{lj} g_{k},
\label{00-vec}
 \\
&& 32 \pi G \rho v_{i} +  8 \partial_{i}\dot\psi  = 2 \nabla^{2} g_{i} 
- \dot{\vartheta} \tilde{\epsilon}_{i}{}^{kl} \nabla^{2} \partial_{k} g_{l} 
\nonumber \\
&-& \tilde{\epsilon}_{i}{}^{kl} \left(\partial_{j} \dot{\vartheta} \right) \left(\partial_{jk} g_{l} \right) 
-  \tilde{\epsilon}^{ljk} \left(\partial_{l} \dot{\vartheta} \right) \left(\partial_{ij} g_{k} \right),
 \\
&& 4 \left(\partial_{l} \vartheta\right) \tilde{\epsilon}_{(i}{}^{lk} \left(\partial_{j)k} \dot{\psi} \right)  =  \left( \partial_{l}\vartheta \right) \tilde{\epsilon}_{(i}{}^{lk} \left(\nabla^{2}\partial_{k} g_{j)} \right) 
\nonumber \\
&+&  \tilde{\epsilon}_{(i}{}^{kl} \ddot{\vartheta} \left(\partial_{j)k} g_{l}\right) 
- 2 \tilde{\epsilon}_{(i}{}^{nl} \left(\partial_{nk} \vartheta \right) \partial_{l[j)} g_{k]},
\ea
where we have used the longitudinal gauge condition. 

Once more, when we apply the CS deformed decomposition of Eq.~\eqref{decomp}, the modified field equations simplify. 
To ${\cal{O}}(\epsilon^{3},\zeta^{0})$, only the temporal-spatial component of these equations survive:
\be
\nabla^{2} g_{i}^{\GR} = 16  \pi G \rho v_{i} +  4 \partial_{i}\dot\psi_{\GR}, 
\ee
which leads to the exterior gravitomagnetic solution of Eq.~\eqref{GR-sol} (see eg.~\cite{lrr-2006-3,Alexander:2007zg,Alexander:2007vt}).

To next order [${\cal{O}}(\epsilon^{3},\zeta)$], the field equations become
\ba
&& \tilde{\epsilon}^{ijk} \left(\partial_{i} \vartheta \right) \nabla^{2} \partial_{j} g_{k}^{\GR} = - \tilde{\epsilon}^{ijk}  \left(\partial_{il} \vartheta \right) \partial_{lj} g_{k}^{\GR},
\label{00-vec2}
 \\
&&  2 \nabla^{2} \chi_{i}^{(g)} =  \dot{\vartheta} \tilde{\epsilon}_{i}{}^{kl} \nabla^{2} \partial_{k} g_{l}^{\GR} 
\nonumber \\\
&+& 
\tilde{\epsilon}_{i}{}^{kl} \left(\partial_{j} \dot{\vartheta} \right) \left(\partial_{jk} g_{l}^{\GR} \right) 
+ \tilde{\epsilon}^{ljk} \left(\partial_{l} \dot{\vartheta} \right) \left(\partial_{ij} g_{k}^{\GR} \right),
\label{0i-vec2}
 \\
&& 4 \left(\partial_{l} \vartheta\right) \tilde{\epsilon}_{(i}{}^{lk} \left(\partial_{j)k} \dot{\psi}_{\GR} \right)  = 
\left( \partial_{l}\vartheta \right) \tilde{\epsilon}_{(i}{}^{lk} \left(\nabla^{2}\partial_{k} g_{j)^{\GR}} \right) 
\nonumber \\
&+&  \tilde{\epsilon}_{(i}{}^{kl} \ddot{\vartheta} \left(\partial_{j)k} g_{l}^{\GR}\right) 
- 2 \tilde{\epsilon}_{(i}{}^{nl} \left(\partial_{nk} \vartheta \right) \partial_{l[j)} g_{k]}^{\GR},
\label{ij-vec2}
\ea
With the choice $\vartheta = \vartheta_{c}$, not all the field equations are automatically satisfied, with 
the temporal-spatial component becoming  
\be
\nabla^{2} \chi_{i}^{(g)} =\frac{\dot{\vartheta}}{2} \tilde{\epsilon}_{i}{}^{kl} \nabla^{2} \partial_{k} g_{l}^{\GR},
\label{0i-last}
\ee
whose solution is given in Eq.~\eqref{frame-drag-pot} by noting that $g_{0i} = g_{i}$. If a non-trivial density distribution is
used, such as the homogeneous sphere in Sec.~\ref{astroph}, then one must ensure that the solution to Eq.~\eqref{0i-last} accounts
for possible boundary contributions that arise to guarantee the junction conditions are satisfied. 

Let us now argue that the conditions of Eq.~\eqref{suff-cond} together with the modified field equations to this order lead 
directly to $\vartheta = \vartheta_{c}$. With either condition, one can
show that Eq.~\eqref{00-vec2} and the second line of Eq.~\eqref{0i-vec2} either automatically vanish or become of ${\cal{O}}(\epsilon^{4},\zeta)$. 
Equation~\eqref{ij-vec2}, on the other hand, leads to $ \tilde{\epsilon}_{(i}{}^{kl} \ddot{\vartheta} \left(\partial_{j)k} g_{l}^{\GR}\right)  = 0$, which
forces $\ddot\vartheta = 0$ since $g_{i}^{\GR}$ does not vanish. Combining Eq.~\eqref{suff-cond} with $\ddot\vartheta =0$ one is then led
to $\vartheta = \vartheta_{c}$. 

Once more, although the conditions in Eq.~\eqref{suff-cond} are sufficient, we have not succeeded in 
formally proving that they are necessary to satisfy the modified field equations. 
In other words, we could not mathematically prove that Eqs.~\eqref{00-vec2}-~\eqref{ij-vec2} does not 
possess some other obscure solution that we have missed. 
 
\subsection{The Pontryagin Constraint}

One last issue to consider is whether the solution found above satisfies the Pontryagin constraint of the non-dynamical
theory. The Pontryagin density is independent of the CS scalar, but for the line element in Eq.~\eqref{metric}, it is given by  
\be
\label{pont-Mink}
\pont = -4 \tilde{\epsilon}^{ijk} \left( \partial_{j}{}^{l} g_{i} \right) \partial_{kl}\left(\phi + \psi\right) + {\cal{O}}(\epsilon^{6}).
\ee
With the GR solutions of Eq.~\eqref{GR-sol}, the Pontryagin density identically vanishes to this order, since 
$\partial_{i} r_{1,2} = n_{1,2}^{i}$ and the Levi-Civita symbol is completely antisymmetric under index permutation. 
If one were to include spin correction to the gravitomagnetic components, then the Pontryagin density would not vanish. In this sense, 
the Pontryagin density, to leading order, is of ${\cal{O}}(\epsilon^{6})$, and thus, unless a post-Newtonian expansion is carried to high
order, this constraint does not affect the analysis of this paper.

\bibliography{phyjabb,master}

\begin{thebibliography}{37}
\expandafter\ifx\csname natexlab\endcsname\relax\def\natexlab#1{#1}\fi
\expandafter\ifx\csname bibnamefont\endcsname\relax
  \def\bibnamefont#1{#1}\fi
\expandafter\ifx\csname bibfnamefont\endcsname\relax
  \def\bibfnamefont#1{#1}\fi
\expandafter\ifx\csname citenamefont\endcsname\relax
  \def\citenamefont#1{#1}\fi
\expandafter\ifx\csname url\endcsname\relax
  \def\url#1{\texttt{#1}}\fi
\expandafter\ifx\csname urlprefix\endcsname\relax\def\urlprefix{URL }\fi
\providecommand{\bibinfo}[2]{#2}
\providecommand{\eprint}[2][]{\url{#2}}

\bibitem[{\citenamefont{Jackiw and Pi}(2003)}]{jackiw:2003:cmo}
\bibinfo{author}{\bibfnamefont{R.}~\bibnamefont{Jackiw}} \bibnamefont{and}
  \bibinfo{author}{\bibfnamefont{S.~Y.} \bibnamefont{Pi}},
  \bibinfo{journal}{Phys. Rev.} \textbf{\bibinfo{volume}{D68}},
  \bibinfo{pages}{104012} (\bibinfo{year}{2003}), \eprint{gr-qc/0308071}.

\bibitem[{\citenamefont{Smith et~al.}(2008)\citenamefont{Smith, Erickcek,
  Caldwell, and Kamionkowski}}]{Smith:2007jm}
\bibinfo{author}{\bibfnamefont{T.~L.} \bibnamefont{Smith}},
  \bibinfo{author}{\bibfnamefont{A.~L.} \bibnamefont{Erickcek}},
  \bibinfo{author}{\bibfnamefont{R.~R.} \bibnamefont{Caldwell}},
  \bibnamefont{and}
  \bibinfo{author}{\bibfnamefont{M.}~\bibnamefont{Kamionkowski}},
  \bibinfo{journal}{Phys. Rev.} \textbf{\bibinfo{volume}{D77}},
  \bibinfo{pages}{024015} (\bibinfo{year}{2008}), \eprint{0708.0001}.

\bibitem[{\citenamefont{Polchinski}(1998)}]{Polchinski:1998rr}
\bibinfo{author}{\bibfnamefont{J.}~\bibnamefont{Polchinski}},
  \emph{\bibinfo{title}{String theory. Vol. 2: Superstring theory and beyond}}
  (\bibinfo{publisher}{Cambridge University Press},
  \bibinfo{address}{Cambridge, UK}, \bibinfo{year}{1998}).

\bibitem[{\citenamefont{Alexander and Gates}(2006)}]{Alexander:2004xd}
\bibinfo{author}{\bibfnamefont{S.~H.~S.} \bibnamefont{Alexander}}
  \bibnamefont{and} \bibinfo{author}{\bibfnamefont{J.}~\bibnamefont{Gates},
  \bibfnamefont{S.~James}}, \bibinfo{journal}{JCAP}
  \textbf{\bibinfo{volume}{0606}}, \bibinfo{pages}{018} (\bibinfo{year}{2006}),
  \eprint{hep-th/0409014}.

\bibitem[{\citenamefont{Svrcek and Witten}(2006)}]{Svrcek:2006yi}
\bibinfo{author}{\bibfnamefont{P.}~\bibnamefont{Svrcek}} \bibnamefont{and}
  \bibinfo{author}{\bibfnamefont{E.}~\bibnamefont{Witten}},
  \bibinfo{journal}{JHEP} \textbf{\bibinfo{volume}{06}}, \bibinfo{pages}{051}
  (\bibinfo{year}{2006}), \eprint{hep-th/0605206}.

\bibitem[{\citenamefont{Alvarez-Gaume and Witten}(1984)}]{AlvarezGaume:1983ig}
\bibinfo{author}{\bibfnamefont{L.}~\bibnamefont{Alvarez-Gaume}}
  \bibnamefont{and} \bibinfo{author}{\bibfnamefont{E.}~\bibnamefont{Witten}},
  \bibinfo{journal}{Nucl. Phys.} \textbf{\bibinfo{volume}{B234}},
  \bibinfo{pages}{269} (\bibinfo{year}{1984}).

\bibitem[{\citenamefont{Grumiller and Yunes}(2008)}]{Grumiller:2007rv}
\bibinfo{author}{\bibfnamefont{D.}~\bibnamefont{Grumiller}} \bibnamefont{and}
  \bibinfo{author}{\bibfnamefont{N.}~\bibnamefont{Yunes}},
  \bibinfo{journal}{Phys. Rev.} \textbf{\bibinfo{volume}{D77}},
  \bibinfo{pages}{044015} (\bibinfo{year}{2008}), \eprint{0711.1868}.

\bibitem[{\citenamefont{Taveras and Yunes}(2008)}]{Taveras:2008yf}
\bibinfo{author}{\bibfnamefont{V.}~\bibnamefont{Taveras}} \bibnamefont{and}
  \bibinfo{author}{\bibfnamefont{N.}~\bibnamefont{Yunes}},
  \bibinfo{journal}{Phys. Rev.} \textbf{\bibinfo{volume}{D78}},
  \bibinfo{pages}{064070} (\bibinfo{year}{2008}), \eprint{0807.2652}.

\bibitem[{\citenamefont{Mercuri and Taveras}(2009)}]{Mercuri:2009zt}
\bibinfo{author}{\bibfnamefont{S.}~\bibnamefont{Mercuri}} \bibnamefont{and}
  \bibinfo{author}{\bibfnamefont{V.}~\bibnamefont{Taveras}}
  (\bibinfo{year}{2009}), \eprint{0903.4407}.

\bibitem[{\citenamefont{Mercuri}(2009{\natexlab{a}})}]{Mercuri:2009vk}
\bibinfo{author}{\bibfnamefont{S.}~\bibnamefont{Mercuri}}
  (\bibinfo{year}{2009}{\natexlab{a}}), \eprint{0903.2270}.

\bibitem[{\citenamefont{Mercuri}(2009{\natexlab{b}})}]{Mercuri:2009zi}
\bibinfo{author}{\bibfnamefont{S.}~\bibnamefont{Mercuri}}
  (\bibinfo{year}{2009}{\natexlab{b}}), \eprint{0902.2764}.

\bibitem[{\citenamefont{Calcagni and Mercuri}(2009)}]{Calcagni:2009xz}
\bibinfo{author}{\bibfnamefont{G.}~\bibnamefont{Calcagni}} \bibnamefont{and}
  \bibinfo{author}{\bibfnamefont{S.}~\bibnamefont{Mercuri}}
  (\bibinfo{year}{2009}), \eprint{0902.0957}.

\bibitem[{\citenamefont{Alexander et~al.}(2007)\citenamefont{Alexander, Peskin,
  and Sheikh-Jabbari}}]{Alexander:2007qe}
\bibinfo{author}{\bibfnamefont{S.~H.} \bibnamefont{Alexander}},
  \bibinfo{author}{\bibfnamefont{M.~E.} \bibnamefont{Peskin}},
  \bibnamefont{and} \bibinfo{author}{\bibfnamefont{M.~M.}
  \bibnamefont{Sheikh-Jabbari}} (\bibinfo{year}{2007}),
  \eprint{hep-ph/0701139}.

\bibitem[{\citenamefont{Alexander and
  Yunes}(2007{\natexlab{a}})}]{Alexander:2007zg}
\bibinfo{author}{\bibfnamefont{S.}~\bibnamefont{Alexander}} \bibnamefont{and}
  \bibinfo{author}{\bibfnamefont{N.}~\bibnamefont{Yunes}},
  \bibinfo{journal}{Phys. Rev. Lett.} \textbf{\bibinfo{volume}{99}},
  \bibinfo{pages}{241101} (\bibinfo{year}{2007}{\natexlab{a}}),
  \eprint{hep-th/0703265}.

\bibitem[{\citenamefont{Alexander and
  Yunes}(2007{\natexlab{b}})}]{Alexander:2007vt}
\bibinfo{author}{\bibfnamefont{S.}~\bibnamefont{Alexander}} \bibnamefont{and}
  \bibinfo{author}{\bibfnamefont{N.}~\bibnamefont{Yunes}},
  \bibinfo{journal}{Phys. Rev.} \textbf{\bibinfo{volume}{D75}},
  \bibinfo{pages}{124022} (\bibinfo{year}{2007}{\natexlab{b}}),
  \eprint{0704.0299}.

\bibitem[{\citenamefont{Alexander et~al.}(2006)\citenamefont{Alexander, Peskin,
  and Sheik-Jabbari}}]{alexander:2004:lfg}
\bibinfo{author}{\bibfnamefont{S.~H.~S.} \bibnamefont{Alexander}},
  \bibinfo{author}{\bibfnamefont{M.~E.} \bibnamefont{Peskin}},
  \bibnamefont{and} \bibinfo{author}{\bibfnamefont{M.~M.}
  \bibnamefont{Sheik-Jabbari}}, \bibinfo{journal}{Phys. Rev. Lett.}
  \textbf{\bibinfo{volume}{96}}, \bibinfo{pages}{081301}
  (\bibinfo{year}{2006}), \eprint{hep-th/0403069}.

\bibitem[{\citenamefont{Konno et~al.}(2008)\citenamefont{Konno, Matsuyama,
  Asano, and Tanda}}]{Konno:2008np}
\bibinfo{author}{\bibfnamefont{K.}~\bibnamefont{Konno}},
  \bibinfo{author}{\bibfnamefont{T.}~\bibnamefont{Matsuyama}},
  \bibinfo{author}{\bibfnamefont{Y.}~\bibnamefont{Asano}}, \bibnamefont{and}
  \bibinfo{author}{\bibfnamefont{S.}~\bibnamefont{Tanda}},
  \bibinfo{journal}{Phys. Rev.} \textbf{\bibinfo{volume}{D78}},
  \bibinfo{pages}{024037} (\bibinfo{year}{2008}), \eprint{0807.0679}.

\bibitem[{\citenamefont{Cantcheff}(2008)}]{Cantcheff:2008qn}
\bibinfo{author}{\bibfnamefont{M.~B.} \bibnamefont{Cantcheff}},
  \bibinfo{journal}{Phys. Rev.} \textbf{\bibinfo{volume}{D78}},
  \bibinfo{pages}{025002} (\bibinfo{year}{2008}), \eprint{0801.0067}.

\bibitem[{\citenamefont{Alexander and Yunes}(2008)}]{Alexander:2008wi}
\bibinfo{author}{\bibfnamefont{S.}~\bibnamefont{Alexander}} \bibnamefont{and}
  \bibinfo{author}{\bibfnamefont{N.}~\bibnamefont{Yunes}},
  \bibinfo{journal}{Phys. Rev.} \textbf{\bibinfo{volume}{D77}},
  \bibinfo{pages}{124040} (\bibinfo{year}{2008}), \eprint{0804.1797}.

\bibitem[{\citenamefont{Alexander et~al.}(2008)\citenamefont{Alexander, Finn,
  and Yunes}}]{Alexander:2007:gwp}
\bibinfo{author}{\bibfnamefont{S.}~\bibnamefont{Alexander}},
  \bibinfo{author}{\bibfnamefont{L.~S.} \bibnamefont{Finn}}, \bibnamefont{and}
  \bibinfo{author}{\bibfnamefont{N.}~\bibnamefont{Yunes}},
  \bibinfo{journal}{Phys. Rev. D} \textbf{\bibinfo{volume}{78}},
  \bibinfo{pages}{066005} (\bibinfo{year}{2008}), \eprint{0712.2542}.

\bibitem[{\citenamefont{Yunes and Finn}(2008)}]{Yunes:2008bu}
\bibinfo{author}{\bibfnamefont{N.}~\bibnamefont{Yunes}} \bibnamefont{and}
  \bibinfo{author}{\bibfnamefont{L.~S.} \bibnamefont{Finn}}
  (\bibinfo{year}{2008}), \eprint{0811.0181}.

\bibitem[{\citenamefont{Konno et~al.}(2007)\citenamefont{Konno, Matsuyama, and
  Tanda}}]{Konno:2007ze}
\bibinfo{author}{\bibfnamefont{K.}~\bibnamefont{Konno}},
  \bibinfo{author}{\bibfnamefont{T.}~\bibnamefont{Matsuyama}},
  \bibnamefont{and} \bibinfo{author}{\bibfnamefont{S.}~\bibnamefont{Tanda}},
  \bibinfo{journal}{Phys. Rev.} \textbf{\bibinfo{volume}{D76}},
  \bibinfo{pages}{024009} (\bibinfo{year}{2007}), \eprint{arXiv:0706.3080
  [gr-qc]}.

\bibitem[{\citenamefont{Tekin}(2007)}]{Tekin:2007rn}
\bibinfo{author}{\bibfnamefont{B.}~\bibnamefont{Tekin}} (\bibinfo{year}{2007}),
  \eprint{arXiv:0710.2528 [gr-qc]}.

\bibitem[{\citenamefont{Guarrera and Hariton}(2007)}]{Guarrera:2007tu}
\bibinfo{author}{\bibfnamefont{D.}~\bibnamefont{Guarrera}} \bibnamefont{and}
  \bibinfo{author}{\bibfnamefont{A.~J.} \bibnamefont{Hariton}}
  (\bibinfo{year}{2007}), \eprint{gr-qc/0702029}.

\bibitem[{\citenamefont{Yunes and Sopuerta}(2008)}]{Yunes:2007ss}
\bibinfo{author}{\bibfnamefont{N.}~\bibnamefont{Yunes}} \bibnamefont{and}
  \bibinfo{author}{\bibfnamefont{C.~F.} \bibnamefont{Sopuerta}},
  \bibinfo{journal}{Phys. Rev.} \textbf{\bibinfo{volume}{D77}},
  \bibinfo{pages}{064007} (\bibinfo{year}{2008}), \eprint{0712.1028}.

\bibitem[{\citenamefont{Burgay et~al.}(2003)}]{Burgay:2003jj}
\bibinfo{author}{\bibfnamefont{M.}~\bibnamefont{Burgay}} \bibnamefont{et~al.},
  \bibinfo{journal}{Nature.} \textbf{\bibinfo{volume}{426}},
  \bibinfo{pages}{531} (\bibinfo{year}{2003}), \eprint{astro-ph/0312071}.

\bibitem[{\citenamefont{Misner et~al.}(1973)\citenamefont{Misner, Thorne, and
  Wheeler}}]{Misner:1973cw}
\bibinfo{author}{\bibfnamefont{C.~W.} \bibnamefont{Misner}},
  \bibinfo{author}{\bibfnamefont{K.}~\bibnamefont{Thorne}}, \bibnamefont{and}
  \bibinfo{author}{\bibfnamefont{J.~A.} \bibnamefont{Wheeler}},
  \emph{\bibinfo{title}{Gravitation}} (\bibinfo{publisher}{W. H. Freeman \&
  Co.}, \bibinfo{address}{San Francisco}, \bibinfo{year}{1973}).

\bibitem[{\citenamefont{Alexander and Yunes}(2007{\natexlab{c}})}]{review}
\bibinfo{author}{\bibfnamefont{S.}~\bibnamefont{Alexander}} \bibnamefont{and}
  \bibinfo{author}{\bibfnamefont{N.}~\bibnamefont{Yunes}}, \bibinfo{journal}{in
  progress}  (\bibinfo{year}{2007}{\natexlab{c}}).

\bibitem[{\citenamefont{Grumiller et~al.}(2008)\citenamefont{Grumiller, Mann,
  and McNees}}]{Grumiller:2008ie}
\bibinfo{author}{\bibfnamefont{D.}~\bibnamefont{Grumiller}},
  \bibinfo{author}{\bibfnamefont{R.}~\bibnamefont{Mann}}, \bibnamefont{and}
  \bibinfo{author}{\bibfnamefont{R.}~\bibnamefont{McNees}},
  \bibinfo{journal}{Phys. Rev.} \textbf{\bibinfo{volume}{D78}},
  \bibinfo{pages}{081502} (\bibinfo{year}{2008}), \eprint{0803.1485}.

\bibitem[{\citenamefont{Yunes and Pretorius}(2009)}]{Yunes:2009hc}
\bibinfo{author}{\bibfnamefont{N.}~\bibnamefont{Yunes}} \bibnamefont{and}
  \bibinfo{author}{\bibfnamefont{F.}~\bibnamefont{Pretorius}}
  (\bibinfo{year}{2009}), \eprint{0902.4669}.

\bibitem[{\citenamefont{Sopuerta and Yunes}(2009)}]{Sopuerta:2009iy}
\bibinfo{author}{\bibfnamefont{C.~F.} \bibnamefont{Sopuerta}} \bibnamefont{and}
  \bibinfo{author}{\bibfnamefont{N.}~\bibnamefont{Yunes}}
  (\bibinfo{year}{2009}), \eprint{0904.4501}.

\bibitem[{\citenamefont{Montenbruck and Gill}(2000)}]{satbook}
\bibinfo{author}{\bibfnamefont{O.}~\bibnamefont{Montenbruck}} \bibnamefont{and}
  \bibinfo{author}{\bibfnamefont{E.}~\bibnamefont{Gill}},
  \emph{\bibinfo{title}{Satellite Orbits. Models, methods, applications.}}
  (\bibinfo{publisher}{Springer}, \bibinfo{address}{Germany},
  \bibinfo{year}{2000}).

\bibitem[{\citenamefont{Will}(2006)}]{lrr-2006-3}
\bibinfo{author}{\bibfnamefont{C.~M.} \bibnamefont{Will}},
  \bibinfo{journal}{Living Reviews in Relativity} \textbf{\bibinfo{volume}{9}}
  (\bibinfo{year}{2006}),
  \urlprefix\url{http://www.livingreviews.org/lrr-2006-3}.

\bibitem[{\citenamefont{Kramer et~al.}(2006)}]{Kramer:2006nb}
\bibinfo{author}{\bibfnamefont{M.}~\bibnamefont{Kramer}} \bibnamefont{et~al.},
  \bibinfo{journal}{Science} \textbf{\bibinfo{volume}{314}},
  \bibinfo{pages}{97} (\bibinfo{year}{2006}), \eprint{astro-ph/0609417}.

\bibitem[{\citenamefont{Iorio}(2008)}]{Iorio:2008xz}
\bibinfo{author}{\bibfnamefont{L.}~\bibnamefont{Iorio}} (\bibinfo{year}{2008}),
  \eprint{0808.0256}.

\bibitem[{\citenamefont{Alexander and Martin}(2005)}]{alexander:2005:bgw}
\bibinfo{author}{\bibfnamefont{S.}~\bibnamefont{Alexander}} \bibnamefont{and}
  \bibinfo{author}{\bibfnamefont{J.}~\bibnamefont{Martin}},
  \bibinfo{journal}{Phys. Rev.} \textbf{\bibinfo{volume}{D71}},
  \bibinfo{pages}{063526} (\bibinfo{year}{2005}), \eprint{hep-th/0410230}.

\bibitem[{\citenamefont{Blanchet}(2006)}]{lrr-2006-4}
\bibinfo{author}{\bibfnamefont{L.}~\bibnamefont{Blanchet}},
  \bibinfo{journal}{Living Reviews in Relativity} \textbf{\bibinfo{volume}{9}}
  (\bibinfo{year}{2006}),
  \urlprefix\url{http://www.livingreviews.org/lrr-2006-4}.

\end{thebibliography}
\end{document}